\documentclass{article}

     \PassOptionsToPackage{numbers, compress}{natbib}

\usepackage{verbatim}
\usepackage{hyperref}
\let\svthefootnote\thefootnote
\newcommand\freefootnote[1]{%
  \let\thefootnote\relax%
  \footnotetext{#1}%
  \let\thefootnote\svthefootnote%
}

\usepackage[preprint]{neurips_2023}

\usepackage[utf8]{inputenc} 
\usepackage[T1]{fontenc}    
\usepackage{hyperref}       
\usepackage{url}            
\usepackage{booktabs}       
\usepackage{amsfonts}       
\usepackage{nicefrac}       
\usepackage{microtype}      
\usepackage{xcolor}         
\usepackage[pdftex]{graphicx}
\usepackage{amsmath}

\title{Accelerated Modelling of Interfaces for Electronic Devices using Graph Neural Networks}

\author{%
  Pratik Brahma$^{1*}$, Krishnakumar Bhattaram$^{1*}$, Sayeef Salahuddin$^{1,2}$ \\
  \\
  $^{1}$ Department of Electrical Engineering and Computer Sciences\\ 
  University of California Berkeley\\
  $^{2}$ Materials Science Division, Lawrence Berkeley National Laboratory\\
  \texttt{\{pratik\_brahma, krishnabhattaram, sayeef\}@berkeley.edu} \\
}

\begin{document}
\freefootnote{* These authors contributed equally to this work}

\maketitle

\begin{abstract}
Modern microelectronic devices are composed of interfaces between a large number of materials, many of which are in amorphous or polycrystalline phases. 
Modeling such non-crystalline materials using first-principles methods such as density functional theory is often numerically intractable. 
Recently, graph neural networks (GNNs) have shown potential to achieve linear complexity with accuracies comparable to ab-initio methods.
Here, we demonstrate the applicability of GNNs to accelerate the atomistic computational pipeline for predicting macroscopic transistor transport characteristics via learning microscopic physical properties.
We generate amorphous heterostructures, specifically the HfO$_{2}$-SiO$_{2}$-Si semiconductor-dielectric transistor gate stack, via GNN predicted atomic forces, and show excellent accuracy in predicting transport characteristics including injection velocity for nanoslab silicon channels.
This work paves the way for faster and more scalable methods to model modern advanced electronic devices via GNNs.

\end{abstract}

\section{Introduction}
\label{Introduction}
\begin{figure}
  \centering
  \includegraphics[width=0.8\linewidth]{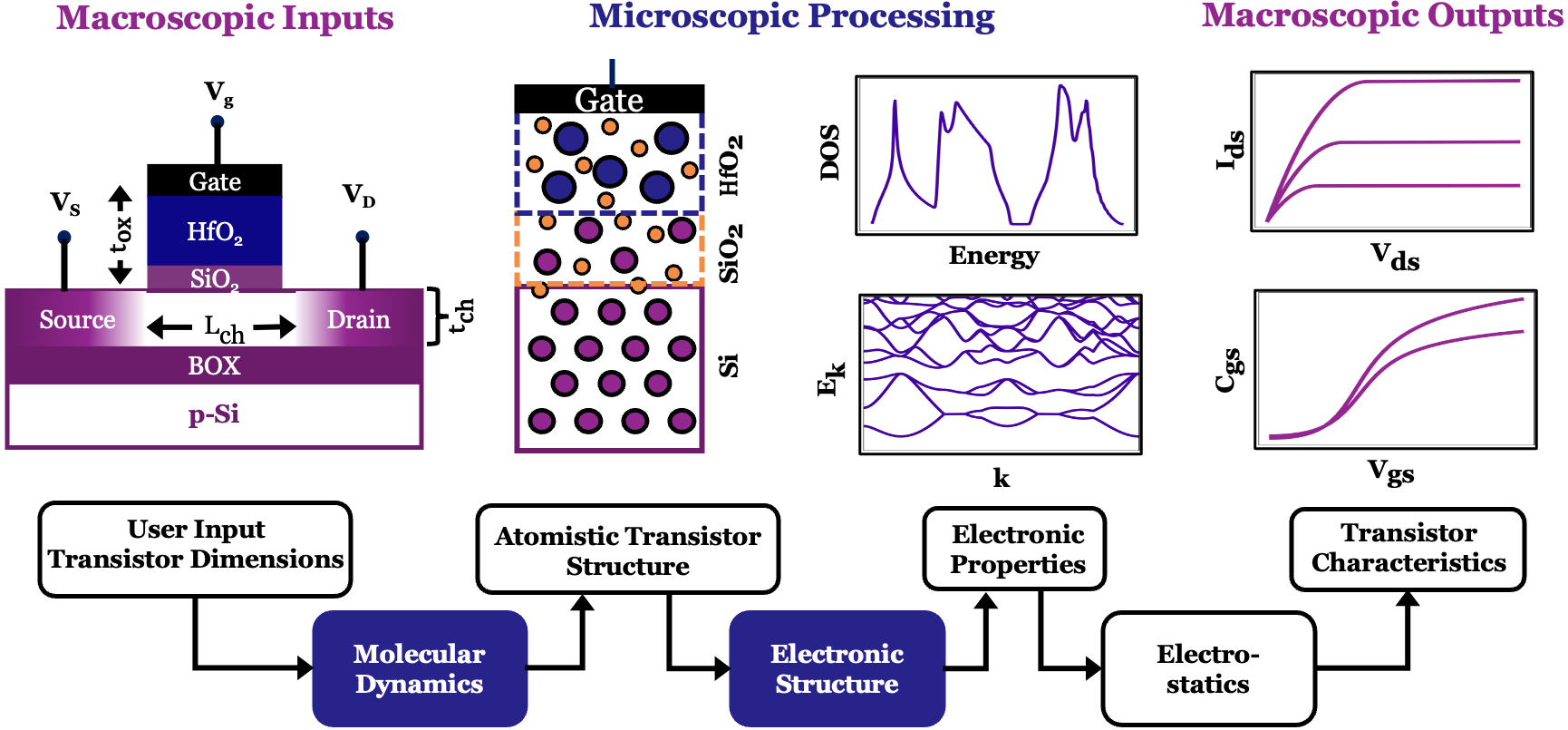}
  \caption{\textbf{Atomistic Computational Pipeline}: Procedure for ab initio accurate predictions of advanced transistor characteristics containing various material interfaces given the macroscopic transistor dimensions. The blue boxes represent the current bottlenecks for scalable simulations of large atomistic devices. We propose to substitute these blocks with GNNs for accelerated predictions.}
  \label{Pipeline}
\end{figure}

The modern Si transistor gate stack comprises of multiple material interfaces whose electronic interactions significantly affect electron transport and ultimately transistor performance. 
In particular, the heterogeneous semiconductor-dielectric gate stack introduces many atomic-scale modeling challenges, which, if addressed, can help design higher-performance gate stacks for next-generation transistors. 
Fundamentally, the starting point for modeling any transport process of an atomistic electronic device stems from the continuity equation $\frac{\partial Q}{\partial t}=-\bar{\nabla}\cdot\bar{J}$, where $Q$ is the non quasi-static charge and $J$ is current, which in turn is a function of $Q$ and injection velocity ($v_{inj}$) \cite{Rahman2003, Natori1994}. 
In practical devices, a significant contribution to $Q$ comes from parasitic sources, which traditional Poisson solvers\cite{Methods_Synopsys_2018} capture well. 
The main challenge in transport calculations is calculating the intrinsic $Q$ and $v_{inj}$, which depend on the specific combination of materials interfaces and confinement effects. 
When amorphous/polycrystalline phases are involved, one cannot directly leverage the E-k diagram.
Therefore, the DOS is then used to calculate all relevant parameters, including $Q$ and $v_{inj}$, which are later used in electrostatic solvers and transport models to directly estimate the fast behavior of nanoscale devices.    
Fig.\ref{Pipeline} summarises this atomistic computational pipeline for calculating transistor characteristics from the macroscopic transistor dimensions.
This pipeline has two bottlenecks for scalability: (i) Molecular Dynamics (MD), which generates atomistic transistor gate stacks with different structural phases, and (ii) Electronic Structure Calculators, which generate atomistic properties like DOS from the quantum Hamiltonian.
These bottlenecks arise as the current state-of-the-art atomistic simulation models\cite{Kresse1996} diagonalize the quantum Hamiltonian, an operation that scales cubically with system size. This poses a challenge for fast and accurate simulations of practically large material systems containing thousands of atoms and varied crystalline states.

Following the success of graph neural networks (GNNs)\cite{Schnet2018, Unke2021} in learning ab initio molecular forces\cite{Park2021} and predicting atomistic properties \cite{Kaundinya2022, Gurunathan2023}, we propose to overcome the scaling challenge by learning the functional relationship between atomic local chemical environments and macroscopic transistor properties. 
As GNN inference (summarized in Fig.\ref{NeuralArchitecture}) scales linearly with system size, orders of magnitude speedup can be realized.
Other existing neural network algorithms for transistor characterization prioritize speed but sacrifice generalizability to unseen geometries and complex material interfaces \cite{Tung2022} by inferring on macroscopic scales and ignoring microscopic properties. 
Our work focuses on learning atomistic contributions to macroscopic transistor characteristics, which we demonstrate yields accurate and generalizable predictions even on unseen transistor geometries.

\section{Methods}
\label{Methods}
\begin{figure}
  \centering
  \includegraphics[width=0.8\linewidth]{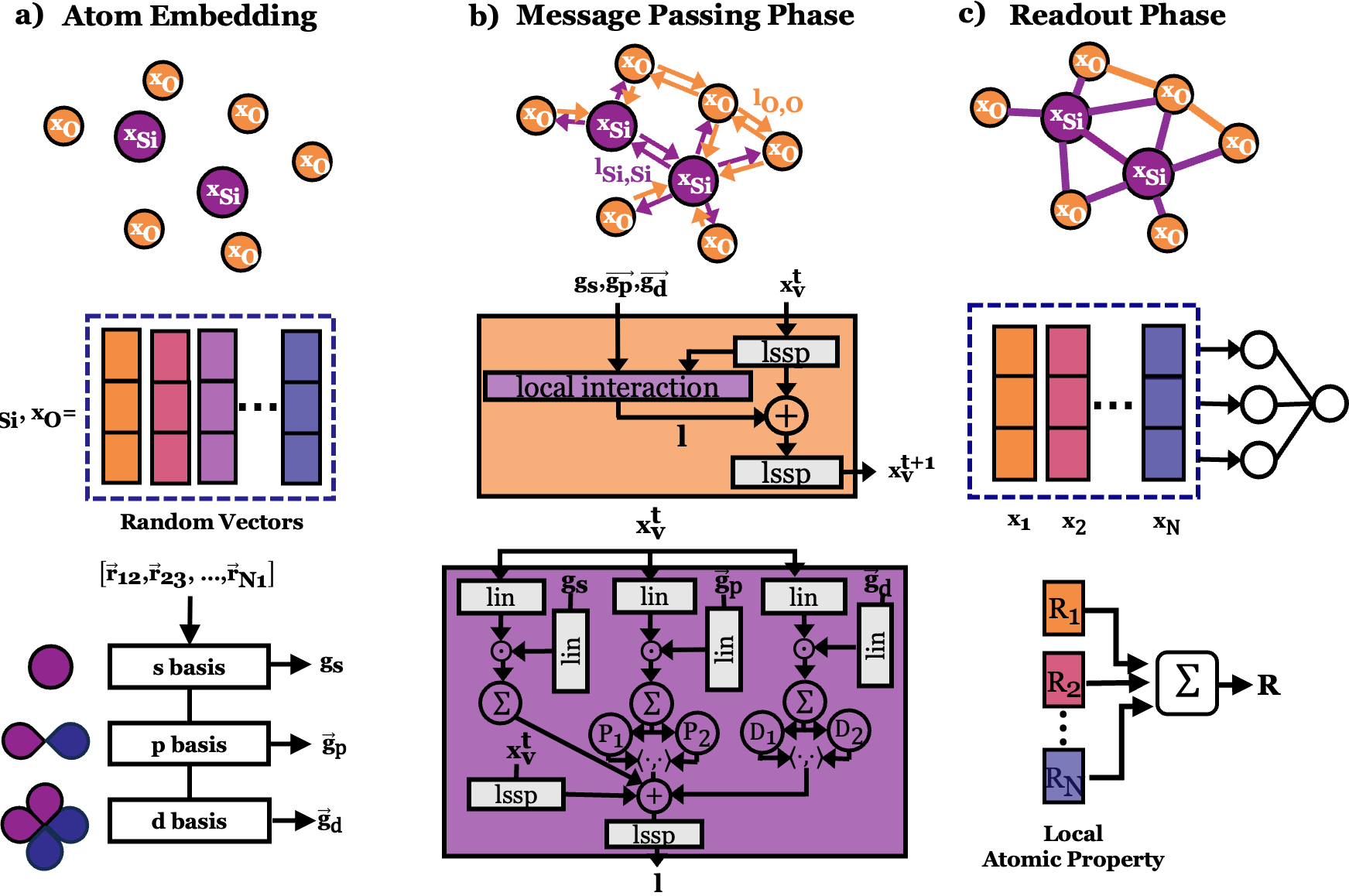}
  \caption{\textbf{Graph Neural Network Architecture}: The forward pass of a GNN is divided into three phases: a) Atom Embedding, b) Message Passing and c) Readout.}
  \label{NeuralArchitecture}
\end{figure}

\paragraph{Neural Network Architecture:} Our GNN architecture is composed of the Schnet\cite{Schnet2018} and SpookyNet\cite{Unke2021} architectures.
The forward pass of the GNN (Fig.\ref{NeuralArchitecture}) is divided into three phases:

\underline{Atom Embedding}: The atomistic transistor structure is modeled as a graph where each atom is a node, and each neighboring atom interaction within a given cutoff $r_c$ is a weighted edge.
Every node is initialized with a random vector ($x_v^t$) according to its atomic number, and every distance vector ($\vec{r}$) to its neighboring atoms is projected in radial ($R$) and spherical harmonic basis ($Y_l^m$)\cite{Unke2021}.
This phase emulates a system of isolated atoms unaware of its local chemical environment. 

\underline{Message Passing}: This phase starts with an isolated system of atoms and allows interactions with atomic neighbors to influence the atomic chemical state.
Continuous convolutional filter-generated messages ($m^t_v$) are sent over the edges as given in Fig.\ref{NeuralArchitecture}b.
The state vector ($x_v^t$) of each node is updated by summing up the incoming messages $(m_{vj}^t)$ along the edges $j$ as $x_v^{t + 1} = x_v^t +  \sum_{j \in \mathcal{N}(v)} m_{vj}^t$.

\underline{Readout}: This phase calculates the local atomic contributions of the desired global property from the state vector of all nodes.
We focus on two sets of properties: (i) Energy and Atomic Forces: These properties generate the atomistic transistor gate stack using MD. 
A dense linear layer predicts the local atomic energy using the final atomic state of each node.
Summing up the local atomistic energy predictions give total energy, and its derivative gives the atomic forces.
(ii) Injection Velocity: This property characterizes the drain current through small channel transistors\cite{Rahman2003, Natori1994}. 
It relates to the average velocity of all electrons over the source-channel potential barrier ($U$). 
In the ballistic limit, the drain current ($I_D$) through a transistor is related to the injection velocity as $I_D = qN_{inv}v_{inj}$, where:
\begin{align*}
        v_{inj} &= \frac{\int dE v_x(E)D(E)f(E + U - E_f)}{N_{inv}}, \quad 
        N_{inv} &= \int dE D(E)f(E + U - E_f)
\end{align*}
$N_{inv}$ is the inversion electron density present in the silicon channel,
$v_x(E)$ is the bandstructure velocity of the electron at energy $E$ and $D(E)$ is the DOS in the silicon channel. 
A dense linear layer with multiple outputs predicts $D(E)$ and $J_x(E) = v_x(E)D(E)$ from the final node state vectors.
We perform PCA on the dataset to reduce the number of output nodes\cite{Bang2021}.
$v_{inj}$ and $N_{inv}$ are subsequently calculated using the predicted properties at a given fermi level ($E_f$).

\paragraph{Datasets:} 
Two datasets are generated related to the transistor gate stack and silicon channel.

\underline{Gate Stack:} The transistor gate stack, a heterostructure of crystalline silicon, amorphous silica, and hafnia, is generated via a high-temperature quench using the LAMMPS MD package\cite{LAMMPS}.
Forces between atoms are estimated using the Charge-Optimized Many Body (COMB) potential\cite{Shan2010, Shan2010_2}.
The crystalline forms $\beta$-cristobalite and orthorhombic HfO$_2$ are melted using constant number, pressure, temperature (NPT) dynamics at 2700K and 3500K, respectively. 
Subsequently, we reshape the melts to match the silicon substrate area using non-equilibrium 
constant number, volume, temperature (NVT) dynamics and quenched via a damped force minimizer.
The generated amorphous silica, hafnia, and crystalline silicon are stacked on each other, and the material interfaces are allowed to relax using the COMB potential. 
This procedure generates a dataset of $\sim$200k molecular structures ranging from 25-96 atoms.

\underline{Silicon channel:} We consider the silicon channel of the transistor as a nanoslab passivated by hydrogen atoms. 
The empirical $sp^3d^5s^*$ tight-binding model in Quantum ATK\cite{smidstrup2020, Liu2008} generates the electronic properties of the silicon channel, primarily $D(E)$ and $J_x(E)$.
Around 1k structures are generated by varying the strain (0.900-1.035) and the nanoslab silicon channel thickness (0.6-2.4 nm).

\section{Results}
\label{Results}
\begin{figure}
  \centering
  \includegraphics[width=0.9\linewidth]{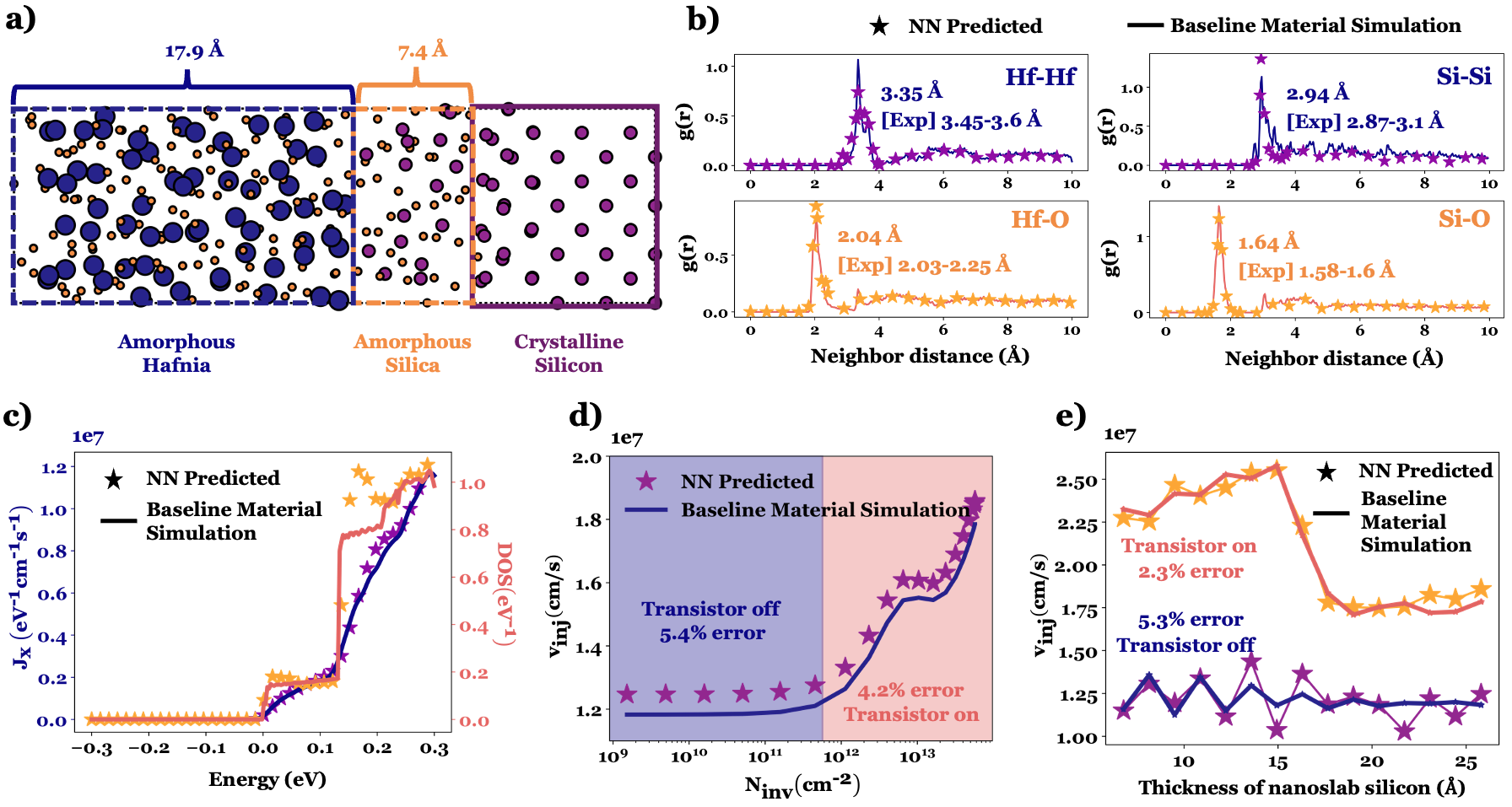}
  \caption{\textbf{Gate Stack Generation and Injection Velocity Prediction}: (a) The gate stack heterostructure generated via the GNN predicted atomic forces. (b) Excellent agreement between the $g(r)$ of the heterostructure generated by GNN and the baseline material simulation. [Exp] refers to the experimental bond length values\cite{ma10111290, ma13194393}. (c), (d) Excellent agreement of the predicted $D(E)$, $J_x(E)$, and $v_{inj}$ between the GNN and the baseline model for a 2.6nm thick unstrained nanoslab silicon. (e) GNN correctly predicts the transistor $v_{inj}$ trend as a function of the unstrained nanoslab silicon channel thickness.}
  \label{Results_Fig}
\end{figure} 

\paragraph{Gate Stack Generation:} We simultaneously train on atomic forces and energy using a 90-10 weighted sum of mean squared error (MSE) losses, which yields a final mean absolute error of 3.0e-2 eV/\AA~for the MD dataset on force predictions. 
The trained model is then used to generate gate stacks of around 200 atoms, a factor of 2 larger than the structures provided during training.
The validity of the GNN-generated heterostructure (Fig.\ref{Results_Fig}a) is confirmed by the excellent match of the pair distribution functions $g(r)$ of amorphous silica and hafnia to the baseline simulation model (Fig.\Ref{Results_Fig}b).
Predicted Si-Si, Hf-Hf, Si-O, and Hf-O bond lengths are within $3\%$ of their experimental values Fig.\Ref{Results_Fig}b)\cite{ma10111290, ma13194393}
This demonstrates that our approach of learning local chemical environments can be generalized to predicting atomic forces and energy in transistor geometries unseen in the initial training set.

\paragraph{Injection Velocity:}
We simultaneously train on $D(E)$ and $J_x(E)$ using an equally weighted sum of MSE losses, which yields a final mean absolute error of 9.0e-4 /eV (0.18\% error) for $D(E)$ and 4.9e4 cm/s-eV (0.82\% error) for $J_x(E)$.
The trained task-specific GNN predicts PCA coefficients for $D(E)$ and $J_x(E)$ over 200 and 15 basis functions respectively for the crystalline silicon nanoslab. 
We found high model performance in reproducing $D(E)$ and $J_x(E)$ for a 2.6nm thick unstrained silicon nanoslab, a structure larger than in the training set, as shown in Fig.\ref{Results_Fig}c. 
From these predictions, we derived $v_{inj}$ for a range of chemical potentials $E_{f}$, finding errors within 5.4\% for both the on and off states of the transistor (Fig. \ref{Results}d)
Furthermore, we evaluate our neural network on unstrained silicon nanoslabs with thicknesses ranging from 0.67 to 2.4 nm. 
We reproduce $v_{inj}$ as a function of thickness with high fidelity (within 5.3\%) (Fig.\ref{Results_Fig}e), demonstrating the ability of our model to successfully predict macroscopic dynamics of unseen geometries of silicon channel.
The runtime for predicting injection velocity by the GNN is around 20ms, while the baseline simulation takes 430s, demonstrating four orders of speed improvement.

\section{Conclusion}
\label{Conclusion}
In this study, we demonstrate the efficacy of GNNs to accelerate an end-to-end simulation framework for predicting the material properties of complex material interfaces. 
Starting from macroscopic transistor dimensions, we use our neural network to predict forces for generating the modern transistor gate stack (HfO$_2$-SiO$_2$-Si) with bond lengths within 3\% of the experimental values. 
We furthur reproduce global electronic ($D(E)$, $J_x(E)$) and transport properties ($v_{inj}$) of crystalline silicon channels.
We show agreement within 5.4\% for injection velocity on unseen geometries and demonstrate high performance on a structure size outside the training domain. 
The scalability and accuracy of our predictions over a wide range of material and transport properties demonstrate the viability of our approach for the modeling of advanced heterogeneous devices, paving the way for rapid design and modeling of next-generation transistors.

\section*{Acknowledgements}
This work is supported by the Defense Advanced Research Projects Agency (DARPA) within the Nanosim Microsystems Exploration Program.

\bibliographystyle{abbrvnat}
\small
\bibliography{main}


\end{document}